\documentclass[3p,twocolumn]{elsarticle}

\usepackage{amssymb}
\usepackage{amsmath}
\usepackage{lineno}
\usepackage{xcolor}
\journal{Chaos Solitons and Fractals}

\begin{document}

%\linenumbers

\begin{frontmatter}

\title{Supporting punishment via taxation in a structured population}

\author[label1,label2]{Hsuan-Wei Lee}
\cortext[mail1]{Email addresses: hwwaynelee@gate.sinica.edu.tw; szolnoki.attila@ek.hun-ren.hu}
\author[label3]{Colin Cleveland}
\author[label4]{and Attila Szolnoki}

\address[label1]{Institute of Sociology, Academia Sinica, Taiwan}
\address[label2]{Department of Physics, National Taiwan University, Taiwan}
\address[label3]{Department of Informatics, King's College London, London, UK}
\address[label4]{Institute of Technical Physics and Materials Science, Centre for Energy Research, P.O. Box 49, H-1525 Budapest, Hungary}

\begin{abstract}
Taxes are an essential and uniformly applied institution for maintaining modern societies. However, the levels of taxation remain an intensive debate topic among citizens. If each citizen contributes to common goals, a minimal tax would be sufficient to cover common expenses. However, this is only achievable at high cooperation level; hence, a larger tax bracket is required. A recent study demonstrated that if an appropriate tax partially covers the punishment of defectors, cooperation can be maintained above a critical level of the multiplication factor, characterizing the synergistic effect of common ventures. Motivated by real-life experiences, we revisited this model by assuming an interactive structure among competitors. All other model elements, including the key parameters characterizing the cost of punishment, fines, and tax level, remain unchanged. The aim was to determine how the spatiality of a population influences the competition of strategies when punishment is partly based on a uniform tax paid by all participants. This extension results in a more subtle system behavior in which different ways of coexistence can be observed, including dynamic pattern formation owing to cyclic dominance among competing strategies. 
\end{abstract}


\end{frontmatter}

\section{Introduction}
\label{intro}

Cooperation is essential in society; however, certain members of smaller or larger communities are tempted to defect for higher individual incomes \cite{hardin_g_s68,sigmund_10}. To avoid the negative consequences of ``rational'' strategy selection, we must identify conditions and incentives that promote the evolution of cooperation \cite{nowak_s06}. As a fundamental theoretical framework for this dilemma, the public goods game is frequently used to model the consequences of individual strategy choices \cite{perc_jrsi13,quan_j_c19,hua_sj_pd23,flores_pre23}. In this game, participants may contribute to a common pool, and the enhanced results of their collective efforts are distributed evenly among all group members, independent of their contributions. Although this theoretical model is simple, it captures the essence of conflict in a broad range of seemingly different problems \cite{quan_j_pa21,wang_cq_c23,duan_yx_csf23,lv_r_csf23}.

As the defector strategy has an obvious advantage over altruistic cooperators, we need incentives to mitigate the gap between their individual payoff values. Rewarding by a bonus to the latter group could be a solution \cite{szolnoki_epl10,cheng_f_amc20,hua_sj_eswa24}, or we may fine the selfish players as a form of penalty \cite{yang_hx_epl20,flores_jtb21,lv_amc22,quan_j_jsm20}. The scientific debate on whether ``the carrot or the stick'' is more effective at controlling a population still has no conclusion; however, in the model presented in this study, we focus on the latter option \cite{gao_sp_pre20,wang_sx_pla21,wang_xj_csf22,sun_xp_amc23,ohdaira_plr23}.

Indeed, punishment can be executed in different ways. The first possibility is that a player punishes a defector partner directly during a face-to-face conflict. In this case, which is frequently referred to as individual or peer punishment, the extra cost of punishment is proportional to the number of defectors near the punisher \cite{brandt_prsb03,helbing_njp10}. Consequently, the fine for a defector is also proportionate: more sticks hit harder than a single stick. An institutional incentive occurs when punishers contribute a fixed amount to a special pool that sanctions possible defectors. In this case, referred to as pool punishment, the defector's fine depends solely on whether a punisher is present, whereas the latter strategy has a fixed extra cost \cite{,ohdaira_srep22,szolnoki_pre11}. 
It is important to emphasize that not all cooperators want to bear the additional cost of punishment while still enjoying its positive consequences. Therefore, ``pure'' cooperators can be considered as second-order free-riders, which redefines the original dilemma in an alternative way. This problem cannot be solved in a well-mixed random system \cite{fowler_n05b}; however, it can be solved automatically in a structured population in which players have limited and practically fixed interactions with their neighbors \cite{helbing_ploscb10}.

In addition to the two major methods, there are several other complicated extensions for sanitizing bad behavior, including 
conditional punishment \cite{szolnoki_jtb13,huang_f_srep18},and
probabilistic cost-sharing \cite{chen_xj_njp14,lv_csf23} or
sampling punishment \cite{xiao_jf_pla23}.
An interesting new method of punishment could occur when ordinary players are not involved. Instead, special punishing actors perform this task, who, in return, are exonerated to contribute to the original joint venture \cite{lee_hw_amc22}. In this “mercenary punishment,” the fee for the incentive is collected from all participants, independent of their strategies, via a tax \cite{shen_y_csf23,han_d_amc23}. This idea was recently extended in a study in which Li {\it et al.} restored the original conflict of second-order free-riding and added a new strategy of defectors that participate in the punishment \cite{li_my_csf22}. Accordingly, besides the traditional pure strategies of the cooperators ($C$) and defectors ($D$), we have two punishing strategies. Here, punishing defectors ($P_D$) refuse to contribute to the original pool; however, they are ready to bear the extra cost of punishment. Moreover, punishing cooperators ($P_C$) also contribute to the original pool of the public goods game. Similar to the original mercenary punishment, punishment strategies are equally supported by funds collected from all group members via a tax mechanism.
The key observation of the study mentioned above, which considered a well-mixed population, was that similar strategies, such as pure and punishing defectors or pure and punishing cooperators, coexist, and a critical synergy value separates their dominance in the parameter region. For low synergy values, the $P_D+D$ solution is detected, which is replaced by $P_C+C$ formation above a threshold. 

As already highlighted, assuming an interaction structure among actors can offer new system behavior. This feature is justified in this study because the collected tax is distributed among specific group members; hence, locality is essential. Furthermore, this aspect cannot be handled by the mean-field calculations applied in the original study. 

Motivated by this expectation, we revisit the model by assuming an interaction structure among group members. All other model elements, including the key parameters characterizing the cost of punishment, fines, and tax level, remain unchanged. Our main aim was to determine how the spatiality of a population influences the competition for strategies when punishment is partly based on a uniform tax paid by all participants. Through simulations, we were able to reveal whether the new condition of interactions is a decisive factor in the system behavior. Our study spans the entire parameter space; hence, we can provide a holistic view of the potential consequences of tax-supported punishment when strategies face the traditional dilemma of contributing to a joint venture or bothering with punishment.

\section{Supporting punishment via tax}
\label{def}

In our spatial population, the players are arranged in a square lattice of size $L \times L$ with periodic boundary conditions. During the Monte Carlo simulations, we applied different system sizes ranging from $L=200$ to $L=1000$ to avoid finite size effects.  Initially, each player at site $x$ is designated as either a pure cooperator ($s_x = C$), pure defector ($s_x = D$), punishing cooperator ($s_x = P_C$), or punishing defector ($s_x = P_D$) with equal probability.

Because of the topology of the interaction graph, the focal player $i$ forms a $G = 5$ size group with the nearest neighbors. According to the traditional public goods game, $C$ and $P_C$ players contribute $c = 1$ to a common pool, and their accumulated contributions are enlarged by a multiplication factor $r$. This enhanced amount is evenly distributed among all group members, independent of their proper strategies. The other two strategies, $D$, and $P_D$, do not contribute to the common pool. To prevent members from defecting, punishing players, such as $P_C$ and $P_D$, make extra efforts and punish members who belong to the defector strategies. This results in an additional $G_P$ cost to bear for punishers, whereas defectors are fined an amount $\beta$ from each punisher partner. To ensure the punishment is less challenging, all group players must pay a fixed tax of $T$ distributed among the abovementioned punisher partners. The corresponding payoff gains from group interactions for these strategies are as follows \cite{li_my_csf22}:

\begin{eqnarray}	
	\label{payoff}
	\Pi_C^g &=& r\frac{N_C+N_{P_C}}{G} - 1 - T \\
	%\Pi_D^g &=& r\frac{N_C+N_{P_C}}{G} - \beta (N_{P_C} + N_{P_D}) - T \nonumber\\
	\Pi_D^g &=& \Pi_{C}^g + 1 - \beta (N_{P_C} + N_{P_D}) \nonumber\\
	\Pi_{P_C}^g &=& \Pi_{C}^g - G_P + \frac{G \cdot T}{N_{P_C} + N_{P_D}} \nonumber\\
	%\Pi_{P_C}^g &=& r\frac{N_C+N_{P_C}}{G} - 1 - T - G_P + \frac{G \cdot T}{N_{P_C} + N_{P_D}} \nonumber\\
	%\Pi_{P_D}^g &=& r\frac{N_{C}+N_{P_C}}{G} - \beta (N_{P_C} + N_{P_D} - 1) - T - G_P + \frac{G \cdot T}{N_{P_C} + N_{P_D}}\,\, \nonumber,
	\Pi_{P_D}^g &=& \Pi_{D}^g - G_P + \beta + \frac{G \cdot T}{N_{P_C} + N_{P_D}} \nonumber \\ \nonumber
\end{eqnarray}
where $N_C$, $N_{P_C}$, and $N_{P_D}$ denote the numbers of cooperators, punishing cooperators, and punishing defectors in the group, respectively. Note that the positive $\beta$ in the last row of Eq. \ref{payoff} indicates that a self-punishment of the $P_D$ player is excluded. 

Because of the structured population, a player belongs to $G$ different groups; hence, the total payoff of a player is calculated by summing all $\Pi^g$ values originating from the different groups in which the player is involved.

During the elementary step, the neighboring $x$ and $y$ players are randomly selected. If their $s_x$ and $s_y$ strategies differ, we calculate their total $\Pi_{s_x}$ and $\Pi_{s_y}$ payoff values. Next, player $y$ imitates the $s_x$ strategy of player $x$ with probability

\begin{equation}
	\label{fermi}
	\Gamma(s_x \to s_y)=1/\{1+\exp[(\Pi_{s_y}-\Pi_{s_x}) /K] \}\,\,,
\end{equation}
where parameter $K$ characterizes the uncertainty in strategy adoption. While $K \to \infty$ represents a completely random decision, the $K \to 0$ limit represents a strongly rational case when only those imitations are possible when the strategy difference is positive. In this study, we used $K=0.1$, which is still in the strong selection (rational with some noise) interval; however, nonzero noise prevents the system from being trapped in a specific solution that represents only a local optimum for the entire system.

\section{Results}
\subsection{Systematic exploration of parameter space}

\begin{figure}
	\centering
	\includegraphics[width=8.0cm]{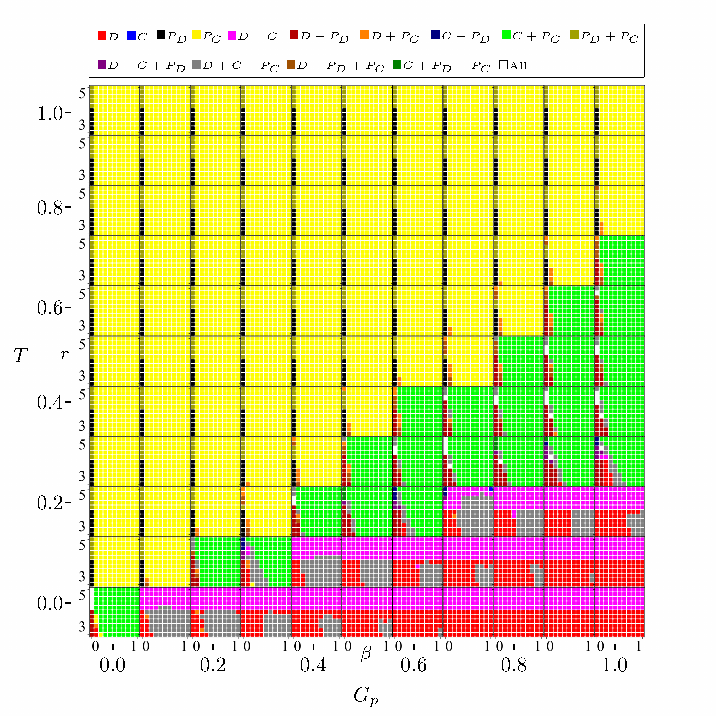}\\
	\caption{Evolutionary solutions on the entire parameter plane of four model parameters. Each subpanel shows the stable solutions on the $\beta-r$ parameter plane at fixed $G_p$ and $T$ values as indicated. The values of $\beta$ vary between 0 and 1, whereas the synergy factor spans the $r \in [3,5]$ interval. The $G_p$ individual cost of punishment and $T$ tax level vary between 0 and 1. The color code of the solutions is shown on the top. The experiments are performed on a linear system size of $L = 400$ grids. \label{global}}
\end{figure}

To gain a comprehensive understanding of the possible system behavior, we systematically scanned all the significant parameters of the proposed model. In particular, we varied the synergy factor, which defines the dilemma strength of the basic conflict, and the tax value, which partly covers punishment. Logically, the individual cost of punishment and the fines change within a reasonable interval. The main observations are summarized in Fig.~\ref{global}, where we present $11^2 = 121$ diagrams of the $r-\beta$ parameter plane for different values of $T$ and $G_P$. Practically, this implies that we reveal the system behavior for $11^4 = 14,641$ combinations of the model parameters. To emphasize the large variety of solutions, we used different colors for the emerging dominant solutions. A color-coded legend is shown at the top of the figure.

In most parameter fields, when tax $T$ exceeds $G_P$, the punishing cooperator strategy can become exclusive and almost independent of the synergy factor. This means that punishing cooperators functions well if the tax value exceeds the individual cost of punishment.
In the other extreme case, when the cost is sufficiently high, and the universal tax is low, pure defectors dominate the entire system. The latter conclusion is valid in the small-$r$ region; otherwise, pure $C$ and $D$ players can coexist owing to the spatial arrangement of their interactions. The winning pair of these nonpunishing strategies is detected at large $r$ values if the tax level is low and is almost independent of the individual $G_P$ cost and $\beta$ fine. 
 
At the intermediate tax level, when the $G_P$ cost is still high, we can identify the $C+P_C$ solution reported earlier for the well-mixed system. Here, the actual fine level has second-order importance.
The alternative pair of $D+P_D$ strategies can also dominate but only within a narrow range of parameter fields. More precisely, we require a small $\beta$ fine at a relatively large individual $G_P$ cost when the synergy factor $r$ is significantly small to achieve this previously reported solution.

Our combined diagrams also reveal several new solutions that have not been previously reported. Evidently, many of them are restricted to a specific combination of model parameters; however, some can be observed in a large area of the parameter space; hence, they can be considered typical system behavior under certain conditions. One of them is a new solution formed by $D+C+P_C$ strategies. This trio can be dominant if $T$ is low, whereas the combination of the remaining three parameters satisfies certain criteria. More precisely, a low fine or large $r$ can be detrimental to this solution for almost all $G_P$ values. Later, we discuss the characteristics of this solution in detail.

\subsection{Competition of solutions in the high cost region}

\begin{figure}
	\centering
	\includegraphics[width=5.5cm]{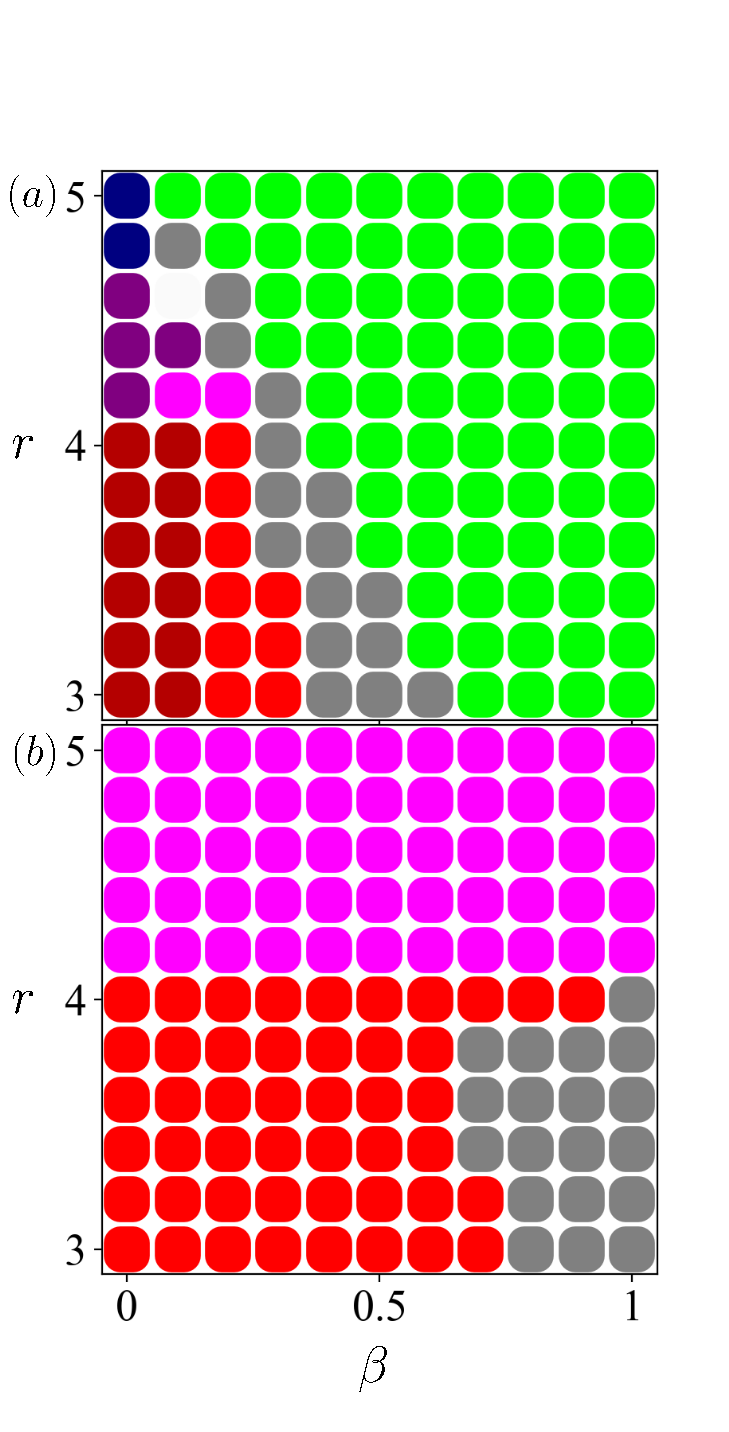}\\
	\caption{Detailed panels of diagrams on $\beta-r$ plane obtained at $G_p=1$, $T=0.3$ (panel~(a) on top) and at $G_p=1$, $T=0.2$ (panel~(b) on bottom). The color code of solutions is identical to those used in Fig.~\ref{global}. The experiments are performed on a linear system size of $L = 1000$ grids.}\label{sub-panels}
\end{figure}

To demonstrate the richness of this model, but highlighting the representative system behavior, we consider an interesting part of the diagrams obtained at high individual cost values. We pick two adjacent whereas very different blocks in Figure ~\ref{global}. Figure~\ref{sub-panels} shows two magnified parts of these diagrams obtained at $G_P=1$, $T=0.3$ (top) and at $G_P=1$, $T=0.2$ (bottom). Although these two cases differ by only $T=0.1$, their stationary solution patterns are quite diverse. Importantly, however, they both represent typical system behaviors because qualitatively similar diagram can also be found for alternative parameter values, as it is shown in Fig.~\ref{global}. The top diagram illustrates that the system behavior could be more delicate than that reported in a random population. At high fine and high $r$ values, the green patches indicate that cooperative strategies form the winning pair. In contrast, at low $\beta$ and low $r$, the defector duo $D+P_D$ prevails, as indicated by dark red patches. Similar system behavior was reported in a random population. However, the other two phases were not revealed earlier. In this study, a full defector state, shown in red, and a three-member solution of $D+C+P_C$ strategy are observed. Gray patches represent the latter.
 
 \begin{figure}
	\centering
	\includegraphics[width=7.5cm]{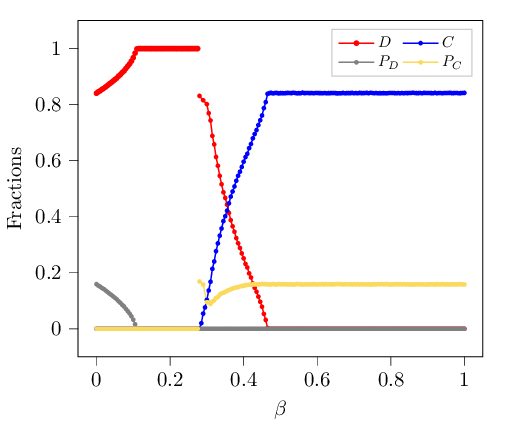}\\
	\caption{Horizontal cross-section of Fig.~\ref{sub-panels}(a) obtained at $r=3.6$, $G_P=1$, and $T=0.3$. Fractions of competing strategies in dependence of punishment fine $\beta$. We can detect three consecutive phase transitions from $D+P_D$ to $D$ to $D+C+P_C$ to $C+P_C$ phases. The second one, when the three-strategy phase replaces the $D$ phase, is discontinuous. The experiments are performed on a linear system size of $L = 1000$ grids.}\label{horizontal}
\end{figure}

The details of these consecutive phase transitions are shown in Fig.~\ref{horizontal}, which is a horizontal cross-section of Fig.~\ref{sub-panels}(a), where we gradually varied the fine level at a fixed $r$ value. Because of the low synergy factor, $r=3.6$, only defector strategies coexist at small $\beta$ values. By increasing this individual fine, $P_D$ players punish each other heavily, resulting in some advantage to pure defectors, whose domain is free from such negative consequences. Consequently, the fraction of $P_D$ players decays gradually, providing space for $D$ players. This full $D$ state is replaced by a three-member solution via a discontinuous phase transition at the critical $\beta_c = 0.27$ fine. By further increasing the level of fine, the defectors die out gradually via a continuous phase transition and enter the $C+P_C$ state. A further increase in $\beta$ has no impact on system behavior in the absence of defector players.

The bottom diagram of Fig.~\ref{sub-panels} shows the system behavior at a slightly lower $T$ tax value, where the individual $G_P$ cost value remains high. Owing to the low tax, pure strategies, which are free from the extra $G_P$ cost, have significant advantages over punishing strategies. Therefore, the four-strategy system is practically reduced to a two-strategy model in which the $\beta$ value is not important, and the boundary separating the full red $D$ and pink $D+C$ solutions is horizontal. However, at high fine values, the homogeneous $D$ state is replaced by an alternative solution, as shown by the gray patches, where cooperation survives.

This phenomenon is illustrated in Fig.~\ref{vertical} where we present a vertical cross-section of the diagram obtained at $\beta=0.9$. As noted, only nonpunishing strategies can survive at high $r$ values. As a well-known effect, the fraction of $C$ players decays gradually with decreasing $r$, and the system finally terminates to a full-defection state. However, the critical $r$, at which this continuous phase transition occurs, is higher than the threshold value obtained in a traditional two-strategy public goods game \cite{szolnoki_pre09c,perc_ejp17}. This deviation can be explained by the absence of universal $T$ tax in the original spatial public goods game. Interestingly, cooperation can re-emerge if the synergy factor decreases further. This solution contains three parties whose specific relationships explain this seemingly paradoxical system behavior. To understand this further, we note that pure cooperators dominate $P_C$ players because of the large $G_P$ additional cost borne by the former strategy. Furthermore, defectors can easily overcome cooperators because of the low synergy factor $r$. Finally, $P_C$ players dominate defectors that suffer from large fine values. Thus, the circle is closed, and the aforementioned strategies can form a cyclic dominant loop at these parameter values. 

\subsection{The emergence of cyclic dominance}

\begin{figure}
	\centering
	\includegraphics[width=7.5cm]{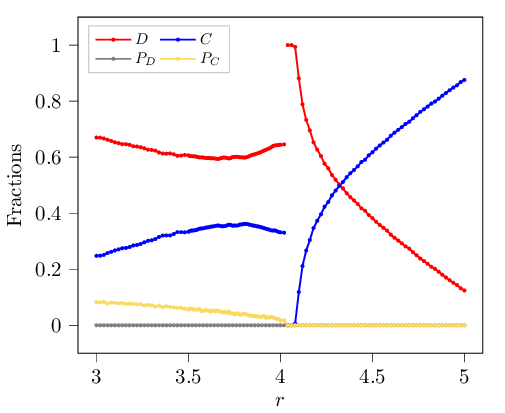}\\
	\caption{Vertical cross-section of Fig.~\ref{sub-panels}(b) obtained at $\beta=0.9$. Fractions of competing strategies in dependence of synergy factor $r$. At high $r$, starting from the ``classic" $D+C$ solution, the system gradually terminates to the full defector state. Interestingly, if we decrease $r$ further, the system enters into a three-strategy state via a discontinuous transition where a significant cooperation level is observed. The experiments are performed on a linear system size of $L = 1000$ grids.}\label{vertical}
\end{figure}

If we check Fig.~\ref{vertical}, we can see that the portion of $P_C$ increases as we decrease the value of $r$. This is again a seemingly paradoxical effect but can be explained as a consequence of intransitive dominance. 
More precisely, it is a well-known fact from the rock-scissors-paper game that if we modify the internal relation between specific members of this loop by increasing the invasion strength of a strategy, then the real beneficial partner will be the one who is the predator of the supported strategy \cite{tainaka_pla95,avelino_pre19b}. Indeed, this is a generally valid observation not restricted to the above-mentioned game; however, it can be detected for social games where competing strategies have similar relations \cite{szolnoki_njp14,szolnoki_csf20b}. To apply this principle to our present case, if we decrease $r$, we directly strengthen defectors; however, this intervention will be utilized by its ``predator'' which is the $P_C$ strategy. Hence, cooperators can survive even at significantly low $r$ values until cyclical dominance is preserved among competitors.

\begin{figure}
	\centering
	\includegraphics[width=7.0cm]{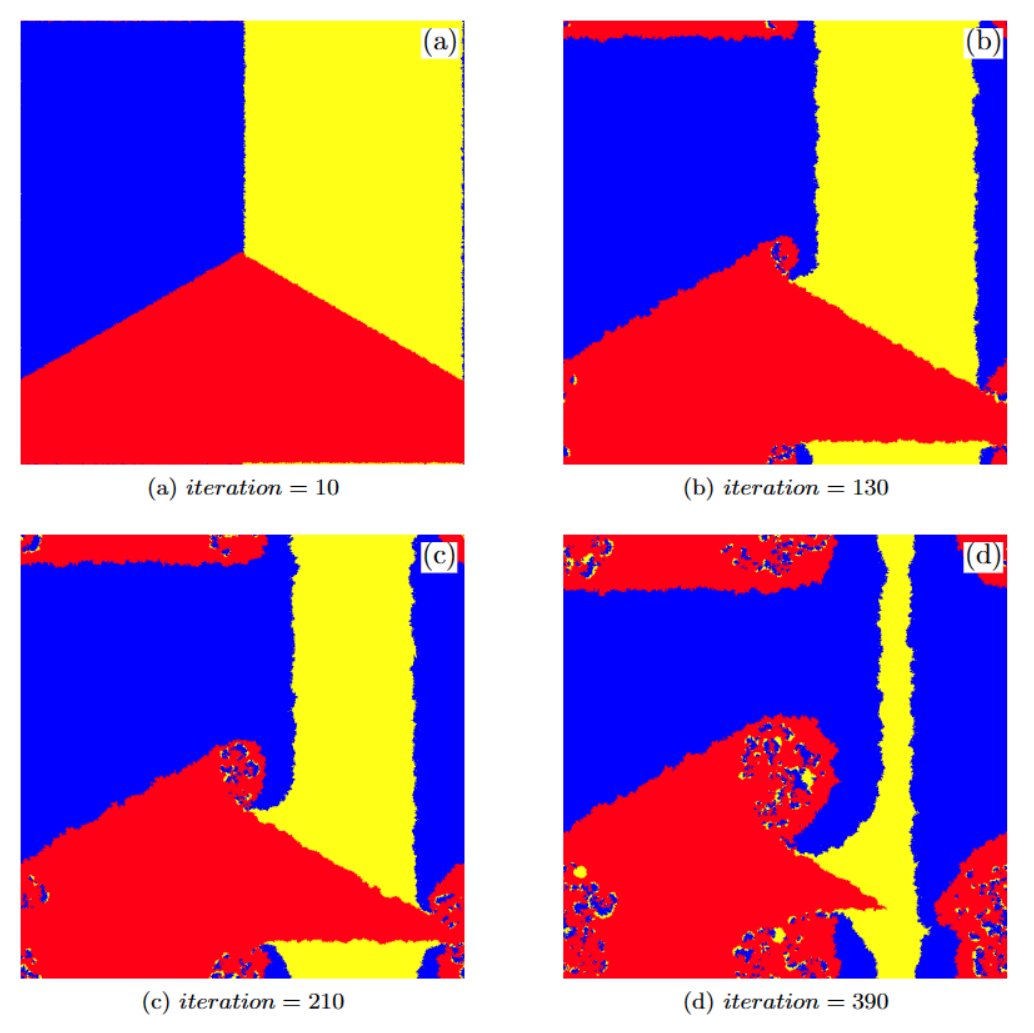}\\
	\caption{Pattern formation owing to cyclic dominance in the $D+C+P_C$ phase. In the initial state, shown in panel~(a), all the strategies form a compact domain. In agreement with Figures~\ref{horizontal}-\ref{vertical}, the red section denotes pure defectors, whereas the blue section represents pure cooperators, and the yellow section represents punishing cooperators. The starting vortex and three anti-vortices at the edges are the origins of rotating spirals, indicating the $D \to C \to P_C \to D$ cyclic dominance among the strategies. Model parameters are $G_P=1$, $T=0.2$, $r=3.2$, $\beta=0.9$. The experiments are performed on a linear system size of $L = 1000$ grids.}\label{pattern}
\end{figure}

To illustrate the intransitive relationship among the strategies mentioned above, we present the evolution of a spatial pattern starting from a specific initial state in Fig.~\ref{pattern}. All three strategies form compact domains, and the vortex at their meeting point has a specific role \cite{szabo_pre99,szolnoki_njp15}. If we launch the evolution, then the frontiers start propagating, and the rotating spirals gradually spread the entire available space. Therefore, the simultaneous presence of all three strategies ensures the stability of the solution \cite{yoshida_srep22,menezes_csf23,avelino_epl21}.
 
\begin{figure}
 	\centering
 	\includegraphics[width=7.5cm]{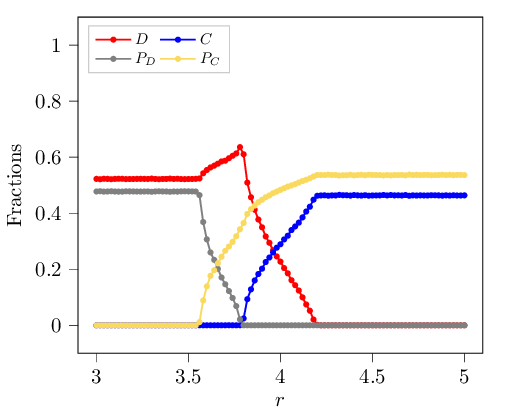}\\
 	\caption{Fractions of strategies in dependence of $r$ obtained at $T=0.4$, $G_P=0.8$, and $\beta=0.1$. The previously reported $D+P_D$ and $C+P_C$ solutions are separated by two new phases where different three-strategy solutions are present. As we increase the synergy factor, the first phase of $D+P_D$ is replaced by the $D+P_D+P_C$ solution, followed by $D+C+P_C$. Lastly, $C+P_C$ dominates the population at high $r$ values. The experiments are performed on a linear system size of $L = 1000$ grids.}\label{two-three}
\end{figure}

To close our study, we stress that the system behavior is even richer and that the spatial setting of players offers several other solutions to emerge. This process is illustrated in Fig.~\ref{global}, where we observe the one-, two-, and three-member solutions. However, in this section, we focused on typical solutions covering most of the parameter space. In addition to the aforementioned states, another three-member solution is briefly discussed. Figure~\ref{two-three} shows a specific example in which two qualitatively different three-member formations separate the previously reported solutions for the defector and cooperator strategy pairs. When the synergy among players is low, the $D$ and $P_D$ strategies coexist. $P_C$ players also become viable and coexist with the aforementioned strategies if we increase $r$. However, this solution differs significantly from the other three strategy formations, which are detected at even higher $r$ values. More precisely, in contrast to the latter solution, no cyclic dominance exists among $D$, $P_D$, and $P_C$ players. Instead, the $P_C$ strategy gradually assumes the role of the $P_D$ strategy for larger $r$ values. The latter players finally die out, which enables $C$ players to establish a stable formation using the remaining two strategies, as discussed above. For higher synergies, only the cooperator strategies remain alive, similar to the situation shown in Fig.~\ref{horizontal}. Note that in contrast to conventional regimes (consisting of only $C$ and $D$), cyclic dominance appears abruptly in Fig.~\ref{horizontal} and vanishes abruptly in Fig.~\ref{vertical}, resulting in discontinuous jumps in their parameter neighborhoods. Conversely, the pattern depicted in Fig.~\ref{two-three} does not apply to the emergence of $P_C$, due to the absence of cyclic dominance among $D$, $P_D$, and $P_C$. $P_C$ is essentially performing the function of $P_D$ during the transition; consequently, the transition is continuous.
 
\section{Conclusions}

The conflict of individual and collective interests is frequently studied in the framework of the public goods game \cite{fu_mj_pa19,zhong_xw_pa21,zheng_jj_pa22,xie_k_csf23,yang_hx_pa19,shen_y_pla22} where the consequence of different conditions and incentives have been checked \cite{xiao_sl_epjb22,kang_hw_pla21,deng_ys_pa21,yang_lh_csf21,feng_s_csf23}. Owing to the large variety of incentives, there is no single answer to how we can support the evolution of cooperation in a particular community formed by humans, animals, or even viruses. However, the situation is complex. Even if we fix an incentive and the related model parameters that determine the individual strategy payoff, system behavior may be sensitive to other conditions, including the potential interactions of competing actors. More precisely, if we abandon the artificial but analytically feasible assumption that players have no stable relations but interact randomly with other population members, then the system behavior could be significantly different from that predicted by a mean-field calculation. Several examples have been previously reported to support this conjecture, and our present study also agrees with these previous studies \cite{liu_jz_epjb21,huang_yc_amc23,yu_fy_csf22,he_jl_pla22,ding_r_csf23}.

Here, our main aim was to check how the spatiality of a population influences the competition for strategies when punishment is partly based on a uniform tax paid by all participants. A previous model study on a random population indicates that the key model parameter is the synergy factor of the public goods game \cite{li_my_csf22}. If it is low, the defector strategies, with or without punishment, dominate and form a stable solution. Above a threshold value of $r$, these actors are replaced by alternative cooperative strategies, where a combination of punishment and taxation can stabilize cooperation. However, our new results reveal a more subtle system behavior in which single-strategy solutions, such as the exclusive dominance of the $P_C$ or $D$ strategy, are also observed. Naturally, the dominant solution could depend sensitively on the actual values of the model parameters; however, we focused on those solutions covering most of the parameter space. Hence, they can be considered representative states of system behavior. Importantly, we observed several more complicated solutions in which three available strategy sets were present. Among these, it is of particular interest when competing strategies cyclically dominate each other, similar to the well-known rock-scissors-paper game \cite{szolnoki_epl20,park_c19b,yang_csf23}. This situation may produce a highly paradoxical system behavior. Therefore, special attention should be paid to modifying the model parameters or changing incentives to reach the desired state of our model. Such a complex system response could be a general lesson for all cases where several competing strategies are present, and their ranks may depend on many parameters.

\vspace{0.5cm}

This research was supported by the National Science and Technology Council of the Republic of China (Taiwan) under Grant NSTC112-2410-H-001-026-MY2 and the National Research, Development and Innovation Office (NKFIH) under Grant No. K142948.

\bibliographystyle{elsarticle-num-names}

\begin{thebibliography}{68}
	\expandafter\ifx\csname natexlab\endcsname\relax\def\natexlab#1{#1}\fi
	\providecommand{\url}[1]{\texttt{#1}}
	\providecommand{\href}[2]{#2}
	\providecommand{\path}[1]{#1}
	\providecommand{\DOIprefix}{doi:}
	\providecommand{\ArXivprefix}{arXiv:}
	\providecommand{\URLprefix}{URL: }
	\providecommand{\Pubmedprefix}{pmid:}
	\providecommand{\doi}[1]{\href{http://dx.doi.org/#1}{\path{#1}}}
	\providecommand{\Pubmed}[1]{\href{pmid:#1}{\path{#1}}}
	\providecommand{\bibinfo}[2]{#2}
	\ifx\xfnm\relax \def\xfnm[#1]{\unskip,\space#1}\fi
	%Type = Article
	\bibitem[{Hardin(1968)}]{hardin_g_s68}
	\bibinfo{author}{G.~Hardin},
	\newblock \bibinfo{title}{The tragedy of the commons},
	\newblock \bibinfo{journal}{Science} \bibinfo{volume}{162}
	(\bibinfo{year}{1968}) \bibinfo{pages}{1243--1248}.
	%Type = Book
	\bibitem[{Sigmund(2010)}]{sigmund_10}
	\bibinfo{author}{K.~Sigmund}, \bibinfo{title}{The Calculus of Selfishness},
	\bibinfo{publisher}{Princeton University Press}, \bibinfo{address}{Princeton,
		NJ}, \bibinfo{year}{2010}.
	%Type = Article
	\bibitem[{Nowak(2006)}]{nowak_s06}
	\bibinfo{author}{M.~A. Nowak},
	\newblock \bibinfo{title}{Five rules for the evolution of cooperation},
	\newblock \bibinfo{journal}{Science} \bibinfo{volume}{314}
	(\bibinfo{year}{2006}) \bibinfo{pages}{1560--1563}.
	%Type = Article
	\bibitem[{Perc et~al.(2013)Perc, G{\'o}mez-Garde{\~n}es, Szolnoki, and
		Flor{\'{\i}a and Y. Moreno}}]{perc_jrsi13}
	\bibinfo{author}{M.~Perc}, \bibinfo{author}{J.~G{\'o}mez-Garde{\~n}es},
	\bibinfo{author}{A.~Szolnoki}, \bibinfo{author}{L.~M. Flor{\'{\i}a and Y.
			Moreno}},
	\newblock \bibinfo{title}{Evolutionary dynamics of group interactions on
		structured populations: a review},
	\newblock \bibinfo{journal}{J. R. Soc. Interface} \bibinfo{volume}{10}
	(\bibinfo{year}{2013}) \bibinfo{pages}{20120997}.
	%Type = Article
	\bibitem[{Quan et~al.(2019)Quan, Li, and Wang}]{quan_j_c19}
	\bibinfo{author}{J.~Quan}, \bibinfo{author}{X.~Li}, \bibinfo{author}{X.~Wang},
	\newblock \bibinfo{title}{The evolution of cooperation in spatial public goods
		game with conditional peer exclusion},
	\newblock \bibinfo{journal}{Chaos} \bibinfo{volume}{29} (\bibinfo{year}{2019})
	\bibinfo{pages}{103137}.
	%Type = Article
	\bibitem[{Hua and Liu(2023)}]{hua_sj_pd23}
	\bibinfo{author}{S.~Hua}, \bibinfo{author}{L.~Liu},
	\newblock \bibinfo{title}{Governance of risky public goods under the threat of
		ostracism},
	\newblock \bibinfo{journal}{Physica D} \bibinfo{volume}{454}
	(\bibinfo{year}{2023}) \bibinfo{pages}{133836}.
	%Type = Article
	\bibitem[{Flores et~al.(2023)Flores, Vainstein, Fernandes, and
		Amaral}]{flores_pre23}
	\bibinfo{author}{L.~S. Flores}, \bibinfo{author}{M.~H. Vainstein},
	\bibinfo{author}{H.~C.~M. Fernandes}, \bibinfo{author}{M.~A. Amaral},
	\newblock \bibinfo{title}{Heterogeneous contributions can jeopardize
		cooperation in the public goods game},
	\newblock \bibinfo{journal}{Phys. Rev. E} \bibinfo{volume}{108}
	(\bibinfo{year}{2023}) \bibinfo{pages}{024111}.
	%Type = Article
	\bibitem[{Quan et~al.(2021)Quan, Tang, and Wang}]{quan_j_pa21}
	\bibinfo{author}{J.~Quan}, \bibinfo{author}{C.~Tang},
	\bibinfo{author}{X.~Wang},
	\newblock \bibinfo{title}{Reputation-based discount effect in imitation on the
		evolution of cooperation in spatial public goods games},
	\newblock \bibinfo{journal}{Physica A} \bibinfo{volume}{563}
	(\bibinfo{year}{2021}) \bibinfo{pages}{125488}.
	%Type = Article
	\bibitem[{Wang and Sun(2023)}]{wang_cq_c23}
	\bibinfo{author}{C.~Wang}, \bibinfo{author}{C.~Sun},
	\newblock \bibinfo{title}{Zealous cooperation does not always promote
		cooperation in public goods games},
	\newblock \bibinfo{journal}{Chaos} \bibinfo{volume}{33} (\bibinfo{year}{2023})
	\bibinfo{pages}{063111}.
	%Type = Article
	\bibitem[{Duan et~al.(2023)Duan, Huang, and Zhang}]{duan_yx_csf23}
	\bibinfo{author}{Y.~Duan}, \bibinfo{author}{J.~Huang},
	\bibinfo{author}{J.~Zhang},
	\newblock \bibinfo{title}{Evolutionary public good games based on the long-term
		payoff mechanism in heterogeneous networks},
	\newblock \bibinfo{journal}{Chaos, Solit. and Fract.} \bibinfo{volume}{174}
	(\bibinfo{year}{2023}) \bibinfo{pages}{113862}.
	%Type = Article
	\bibitem[{Lv et~al.(2023)Lv, Qian, Hao, Wu, Guo, and Ling}]{lv_r_csf23}
	\bibinfo{author}{R.~Lv}, \bibinfo{author}{J.-L. Qian}, \bibinfo{author}{Q.-Y.
		Hao}, \bibinfo{author}{C.-Y. Wu}, \bibinfo{author}{N.~Guo},
	\bibinfo{author}{X.~Ling},
	\newblock \bibinfo{title}{The impact of current and historical reputation with
		non-uniform change on cooperation in spatial public goods game},
	\newblock \bibinfo{journal}{Chaos, Solit. and Fract.} \bibinfo{volume}{175}
	(\bibinfo{year}{2023}) \bibinfo{pages}{113968}.
	%Type = Article
	\bibitem[{Szolnoki and Perc(2010)}]{szolnoki_epl10}
	\bibinfo{author}{A.~Szolnoki}, \bibinfo{author}{M.~Perc},
	\newblock \bibinfo{title}{Reward and cooperation in the spatial public goods
		game},
	\newblock \bibinfo{journal}{EPL} \bibinfo{volume}{92} (\bibinfo{year}{2010})
	\bibinfo{pages}{38003}.
	%Type = Article
	\bibitem[{Cheng et~al.(2020)Cheng, Chen, and Chen}]{cheng_f_amc20}
	\bibinfo{author}{F.~Cheng}, \bibinfo{author}{T.~Chen},
	\bibinfo{author}{Q.~Chen},
	\newblock \bibinfo{title}{Rewards based on public loyalty program promote
		cooperation in public goods game},
	\newblock \bibinfo{journal}{Appl. Math. and Comput.} \bibinfo{volume}{378}
	(\bibinfo{year}{2020}) \bibinfo{pages}{125180}.
	%Type = Article
	\bibitem[{Hua and Liu(2024)}]{hua_sj_eswa24}
	\bibinfo{author}{S.~Hua}, \bibinfo{author}{L.~Liu},
	\newblock \bibinfo{title}{Coevolutionary dynamics of population and
		institutional rewards in public goods games},
	\newblock \bibinfo{journal}{Expert Systems With Applications}
	\bibinfo{volume}{237} (\bibinfo{year}{2024}) \bibinfo{pages}{121579}.
	%Type = Article
	\bibitem[{Yang and Fu(2020)}]{yang_hx_epl20}
	\bibinfo{author}{H.-X. Yang}, \bibinfo{author}{M.-J. Fu},
	\newblock \bibinfo{title}{A punishment mechanism in the spatial public goods
		game with continuous strategies},
	\newblock \bibinfo{journal}{EPL} \bibinfo{volume}{132} (\bibinfo{year}{2020})
	\bibinfo{pages}{10007}.
	%Type = Article
	\bibitem[{Flores et~al.(2021)Flores, Fernandes, Amaral, and
		Vainstein}]{flores_jtb21}
	\bibinfo{author}{L.~S. Flores}, \bibinfo{author}{H.~C. Fernandes},
	\bibinfo{author}{M.~A. Amaral}, \bibinfo{author}{M.~H. Vainstein},
	\newblock \bibinfo{title}{Symbiotic behaviour in the public goods game with
		altruistic punishment},
	\newblock \bibinfo{journal}{J. Theor. Biol.} \bibinfo{volume}{524}
	(\bibinfo{year}{2021}) \bibinfo{pages}{110737}.
	%Type = Article
	\bibitem[{Lv and Song(2022)}]{lv_amc22}
	\bibinfo{author}{S.~Lv}, \bibinfo{author}{F.~Song},
	\newblock \bibinfo{title}{Particle swarm intelligence and the evolution of
		cooperation in the spatial public goods game with punishment},
	\newblock \bibinfo{journal}{Appl. Math. and Comput.} \bibinfo{volume}{412}
	(\bibinfo{year}{2022}) \bibinfo{pages}{126586}.
	%Type = Article
	\bibitem[{Quan et~al.(2020)Quan, Qin, Zhou, Wang, and Yang}]{quan_j_jsm20}
	\bibinfo{author}{J.~Quan}, \bibinfo{author}{Y.~Qin}, \bibinfo{author}{Y.~Zhou},
	\bibinfo{author}{X.~Wang}, \bibinfo{author}{J.-B. Yang},
	\newblock \bibinfo{title}{How to evaluate one's behavior toward 'bad'
		individuals? exploring good social norms in promoting cooperation in spatial
		public goods games},
	\newblock \bibinfo{journal}{J. Stat. Mech.} \bibinfo{volume}{2020}
	(\bibinfo{year}{2020}) \bibinfo{pages}{093405}.
	%Type = Article
	\bibitem[{Gao et~al.(2020)Gao, Du, and Liang}]{gao_sp_pre20}
	\bibinfo{author}{S.~Gao}, \bibinfo{author}{J.~Du}, \bibinfo{author}{J.~Liang},
	\newblock \bibinfo{title}{Evolution of cooperation under punishment},
	\newblock \bibinfo{journal}{Phys. Rev. E} \bibinfo{volume}{101}
	(\bibinfo{year}{2020}) \bibinfo{pages}{062419}.
	%Type = Article
	\bibitem[{Wang et~al.(2021)Wang, Liu, and Chen}]{wang_sx_pla21}
	\bibinfo{author}{S.~Wang}, \bibinfo{author}{L.~Liu}, \bibinfo{author}{X.~Chen},
	\newblock \bibinfo{title}{Tax-based pure punishment and reward in the public
		goods game},
	\newblock \bibinfo{journal}{Phys. Lett. A} \bibinfo{volume}{386}
	(\bibinfo{year}{2021}) \bibinfo{pages}{126965}.
	%Type = Article
	\bibitem[{Wang et~al.(2022)Wang, Ding, Zhao, Chen, and Gu}]{wang_xj_csf22}
	\bibinfo{author}{X.~Wang}, \bibinfo{author}{R.~Ding},
	\bibinfo{author}{J.~Zhao}, \bibinfo{author}{W.~Chen},
	\bibinfo{author}{C.~Gu},
	\newblock \bibinfo{title}{Competition of punishment and reward among
		inequity-averse individuals in spatial public goods games},
	\newblock \bibinfo{journal}{Chaos, Solit. and Fract.}
	\bibinfo{volume}{156} (\bibinfo{year}{2022}) \bibinfo{pages}{111862}.
	%Type = Article
	\bibitem[{Sun et~al.(2023)Sun, Li, Kang, Shen, and Chen}]{sun_xp_amc23}
	\bibinfo{author}{X.~Sun}, \bibinfo{author}{M.~Li}, \bibinfo{author}{H.~Kang},
	\bibinfo{author}{Y.~Shen}, \bibinfo{author}{Q.~Chen},
	\newblock \bibinfo{title}{Combined effect of pure punishment and reward in the
		public goods game},
	\newblock \bibinfo{journal}{Appl. Math. and Comput.} \bibinfo{volume}{445}
	(\bibinfo{year}{2023}) \bibinfo{pages}{127853}.
	%Type = Article
	\bibitem[{Ohdaira(2023)}]{ohdaira_plr23}
	\bibinfo{author}{T.~Ohdaira},
	\newblock \bibinfo{title}{How can we relax the cost of reward and punishment?},
	\newblock \bibinfo{journal}{Phys. Life. Rev.} \bibinfo{volume}{46}
	(\bibinfo{year}{2023}) \bibinfo{pages}{129--130}.
	%Type = Article
	\bibitem[{Brandt et~al.(2003)Brandt, Hauert, and Sigmund}]{brandt_prsb03}
	\bibinfo{author}{H.~Brandt}, \bibinfo{author}{C.~Hauert},
	\bibinfo{author}{K.~Sigmund},
	\newblock \bibinfo{title}{Punishing and reputation in spatial public goods
		games},
	\newblock \bibinfo{journal}{Proc. R. Soc. Lond. Ser B} \bibinfo{volume}{270}
	(\bibinfo{year}{2003}) \bibinfo{pages}{1099--1104}.
	%Type = Article
	\bibitem[{Helbing et~al.(2010)Helbing, Szolnoki, Perc, and
		Szab{\'o}}]{helbing_njp10}
	\bibinfo{author}{D.~Helbing}, \bibinfo{author}{A.~Szolnoki},
	\bibinfo{author}{M.~Perc}, \bibinfo{author}{G.~Szab{\'o}},
	\newblock \bibinfo{title}{Punish, but not too hard: how costly punishment
		spreads in the spatial public goods game},
	\newblock \bibinfo{journal}{New J. Phys.} \bibinfo{volume}{12}
	(\bibinfo{year}{2010}) \bibinfo{pages}{083005}.
	%Type = Article
	\bibitem[{Ohdaira(2022)}]{ohdaira_srep22}
	\bibinfo{author}{T.~Ohdaira},
	\newblock \bibinfo{title}{The probabilistic pool punishment proportional to the
		difference of payoff outperforms previous pool and peer punishment},
	\newblock \bibinfo{journal}{Sci. Rep.} \bibinfo{volume}{12}
	(\bibinfo{year}{2022}) \bibinfo{pages}{6604}.
	%Type = Article
	\bibitem[{Szolnoki et~al.(2011)Szolnoki, Szab{\'o}, and Perc}]{szolnoki_pre11}
	\bibinfo{author}{A.~Szolnoki}, \bibinfo{author}{G.~Szab{\'o}},
	\bibinfo{author}{M.~Perc},
	\newblock \bibinfo{title}{Phase diagrams for the spatial public goods game with
		pool punishment},
	\newblock \bibinfo{journal}{Phys. Rev. E} \bibinfo{volume}{83}
	(\bibinfo{year}{2011}) \bibinfo{pages}{036101}.
	%Type = Article
	\bibitem[{Fowler(2005)}]{fowler_n05b}
	\bibinfo{author}{J.~H. Fowler},
	\newblock \bibinfo{title}{Second-order free-riding problem solved?},
	\newblock \bibinfo{journal}{Nature} \bibinfo{volume}{437}
	(\bibinfo{year}{2005}) \bibinfo{pages}{E8--E8}.
	%Type = Article
	\bibitem[{Helbing et~al.(2010)Helbing, Szolnoki, Perc, and
		Szab{\'o}}]{helbing_ploscb10}
	\bibinfo{author}{D.~Helbing}, \bibinfo{author}{A.~Szolnoki},
	\bibinfo{author}{M.~Perc}, \bibinfo{author}{G.~Szab{\'o}},
	\newblock \bibinfo{title}{Evolutionary establishment of moral and double moral
		standards through spatial interactions},
	\newblock \bibinfo{journal}{PLoS Comput. Biol.} \bibinfo{volume}{6}
	(\bibinfo{year}{2010}) \bibinfo{pages}{e1000758}.
	%Type = Article
	\bibitem[{Szolnoki and Perc(2013)}]{szolnoki_jtb13}
	\bibinfo{author}{A.~Szolnoki}, \bibinfo{author}{M.~Perc},
	\newblock \bibinfo{title}{Effectiveness of conditional punishment for the
		evolution of public cooperation},
	\newblock \bibinfo{journal}{J. Theor. Biol.} \bibinfo{volume}{325}
	(\bibinfo{year}{2013}) \bibinfo{pages}{34--41}.
	%Type = Article
	\bibitem[{Huang et~al.(2018)Huang, Chen, and Wang}]{huang_f_srep18}
	\bibinfo{author}{F.~Huang}, \bibinfo{author}{X.~Chen},
	\bibinfo{author}{L.~Wang},
	\newblock \bibinfo{title}{Conditional punishment is a double-edged sword in
		promoting cooperation},
	\newblock \bibinfo{journal}{Sci. Rep.} \bibinfo{volume}{8}
	(\bibinfo{year}{2018}) \bibinfo{pages}{528}.
	%Type = Article
	\bibitem[{Chen et~al.(2014)Chen, Szolnoki, and Perc}]{chen_xj_njp14}
	\bibinfo{author}{X.~Chen}, \bibinfo{author}{A.~Szolnoki},
	\bibinfo{author}{M.~Perc},
	\newblock \bibinfo{title}{Probabilistic sharing solves the problem of costly
		punishment},
	\newblock \bibinfo{journal}{New J. Phys.} \bibinfo{volume}{16}
	(\bibinfo{year}{2014}) \bibinfo{pages}{083016}.
	%Type = Article
	\bibitem[{Lv et~al.(2023)Lv, Li, and Zhao}]{lv_csf23}
	\bibinfo{author}{S.~Lv}, \bibinfo{author}{J.~Li}, \bibinfo{author}{C.~Zhao},
	\newblock \bibinfo{title}{The evolution of cooperation in voluntary public
		goods game with shared-punishment},
	\newblock \bibinfo{journal}{Chaos, Solit. and Fract.} \bibinfo{volume}{172}
	(\bibinfo{year}{2023}) \bibinfo{pages}{113552}.
	%Type = Article
	\bibitem[{Xiao et~al.(2023)Xiao, Liu, Chen, and Szolnoki}]{xiao_jf_pla23}
	\bibinfo{author}{J.~Xiao}, \bibinfo{author}{L.~Liu}, \bibinfo{author}{X.~Chen},
	\bibinfo{author}{A.~Szolnoki},
	\newblock \bibinfo{title}{Evolution of cooperation driven by sampling
		punishment},
	\newblock \bibinfo{journal}{Phys. Lett. A} \bibinfo{volume}{475}
	(\bibinfo{year}{2023}) \bibinfo{pages}{128879}.
	%Type = Article
	\bibitem[{Lee et~al.(2022)Lee, Cleveland, and Szolnoki}]{lee_hw_amc22}
	\bibinfo{author}{H.-W. Lee}, \bibinfo{author}{C.~Cleveland},
	\bibinfo{author}{A.~Szolnoki},
	\newblock \bibinfo{title}{Mercenary punishment in structured populations},
	\newblock \bibinfo{journal}{Appl. Math. and Comput.}
	\bibinfo{volume}{417} (\bibinfo{year}{2022}) \bibinfo{pages}{126797}.
	%Type = Article
	\bibitem[{Shen et~al.(2023)Shen, Lei, Kang, Li, Sun, and Chen}]{shen_y_csf23}
	\bibinfo{author}{Y.~Shen}, \bibinfo{author}{W.~Lei}, \bibinfo{author}{H.~Kang},
	\bibinfo{author}{M.~Li}, \bibinfo{author}{X.~Sun}, \bibinfo{author}{Q.~Chen},
	\newblock \bibinfo{title}{Evolutionary dynamics of public goods game with
		tax-based rewarding cooperators},
	\newblock \bibinfo{journal}{Chaos, Solit. and Fract.} \bibinfo{volume}{175}
	(\bibinfo{year}{2023}) \bibinfo{pages}{114030}.
	%Type = Article
	\bibitem[{Han and He(2023)}]{han_d_amc23}
	\bibinfo{author}{D.~Han}, \bibinfo{author}{Y.~He},
	\newblock \bibinfo{title}{The impact of labor subsidy, taxation and corruption
		on individual behavior},
	\newblock \bibinfo{journal}{Appl. Math. and Comput.} \bibinfo{volume}{458}
	(\bibinfo{year}{2023}) \bibinfo{pages}{128247}.
	%Type = Article
	\bibitem[{Li et~al.(2022)Li, Kang, Sun, Shen, and Chen}]{li_my_csf22}
	\bibinfo{author}{M.~Li}, \bibinfo{author}{H.~Kang}, \bibinfo{author}{X.~Sun},
	\bibinfo{author}{Y.~Shen}, \bibinfo{author}{Q.~Chen},
	\newblock \bibinfo{title}{Replicator dynamics of public goods game with
		tax-based punishment},
	\newblock \bibinfo{journal}{Chaos, Solit. and Fract.} \bibinfo{volume}{164}
	(\bibinfo{year}{2022}) \bibinfo{pages}{112747}.
	%Type = Article
	\bibitem[{Szolnoki et~al.(2009)Szolnoki, Perc, and Szab{\'o}}]{szolnoki_pre09c}
	\bibinfo{author}{A.~Szolnoki}, \bibinfo{author}{M.~Perc},
	\bibinfo{author}{G.~Szab{\'o}},
	\newblock \bibinfo{title}{Topology-independent impact of noise on cooperation
		in spatial public goods games},
	\newblock \bibinfo{journal}{Phys. Rev. E} \bibinfo{volume}{80}
	(\bibinfo{year}{2009}) \bibinfo{pages}{056109}.
	%Type = Article
	\bibitem[{Perc(2017)}]{perc_ejp17}
	\bibinfo{author}{M.~Perc},
	\newblock \bibinfo{title}{High-performance parallel computing in the classroom
		using the public goods game as an example},
	\newblock \bibinfo{journal}{Eur. J. Phys.} \bibinfo{volume}{38}
	(\bibinfo{year}{2017}) \bibinfo{pages}{045801}.
	%Type = Article
	\bibitem[{Tainaka(1995)}]{tainaka_pla95}
	\bibinfo{author}{K.~Tainaka},
	\newblock \bibinfo{title}{Indirect effect in cyclic voter models},
	\newblock \bibinfo{journal}{Phys. Lett. A} \bibinfo{volume}{207}
	(\bibinfo{year}{1995}) \bibinfo{pages}{53--57}.
	%Type = Article
	\bibitem[{Avelino et~al.(2019)Avelino, \protect{de Oliveira}, and
		Trintin}]{avelino_pre19b}
	\bibinfo{author}{P.~P. Avelino}, \bibinfo{author}{B.~F. \protect{de Oliveira}},
	\bibinfo{author}{R.~S. Trintin},
	\newblock \bibinfo{title}{Predominance of the weakest species in
		\protect{Lotka-Volterra} and \protect{May-Leonard} formulations of the
		rock-paper-scissors model},
	\newblock \bibinfo{journal}{Phys. Rev. E} \bibinfo{volume}{100}
	(\bibinfo{year}{2019}) \bibinfo{pages}{042209}.
	%Type = Article
	\bibitem[{Szolnoki and Perc(2014)}]{szolnoki_njp14}
	\bibinfo{author}{A.~Szolnoki}, \bibinfo{author}{M.~Perc},
	\newblock \bibinfo{title}{Costly hide and seek pays: unexpected consequences of
		deceit in a social dilemma},
	\newblock \bibinfo{journal}{New J. Phys.} \bibinfo{volume}{16}
	(\bibinfo{year}{2014}) \bibinfo{pages}{113003}.
	%Type = Article
	\bibitem[{Szolnoki and Chen(2020)}]{szolnoki_csf20b}
	\bibinfo{author}{A.~Szolnoki}, \bibinfo{author}{X.~Chen},
	\newblock \bibinfo{title}{Strategy dependent learning activity in cyclic
		dominant systems},
	\newblock \bibinfo{journal}{Chaos, Solit. and Fract.} \bibinfo{volume}{138}
	(\bibinfo{year}{2020}) \bibinfo{pages}{109935}.
	%Type = Article
	\bibitem[{Szab{\'o} et~al.(1999)Szab{\'o}, Santos, and Mendes}]{szabo_pre99}
	\bibinfo{author}{G.~Szab{\'o}}, \bibinfo{author}{M.~A. Santos},
	\bibinfo{author}{J.~F.~F. Mendes},
	\newblock \bibinfo{title}{Vortex dynamics in a three-state model under cyclic
		dominance},
	\newblock \bibinfo{journal}{Phys. Rev. E} \bibinfo{volume}{60}
	(\bibinfo{year}{1999}) \bibinfo{pages}{3776--3780}.
	%Type = Article
	\bibitem[{Szolnoki and Perc(2015)}]{szolnoki_njp15}
	\bibinfo{author}{A.~Szolnoki}, \bibinfo{author}{M.~Perc},
	\newblock \bibinfo{title}{Vortices determine the dynamics of biodiversity in
		cyclical interactions with protection spillovers},
	\newblock \bibinfo{journal}{New J. Phys.} \bibinfo{volume}{17}
	(\bibinfo{year}{2015}) \bibinfo{pages}{113033}.
	%Type = Article
	\bibitem[{Yoshida et~al.(2022)Yoshida, Mizoguchi, and
		Hatsugai}]{yoshida_srep22}
	\bibinfo{author}{T.~Yoshida}, \bibinfo{author}{T.~Mizoguchi},
	\bibinfo{author}{Y.~Hatsugai},
	\newblock \bibinfo{title}{\protect{Non‐Hermitian} topology in
		rock-paper-scissors games},
	\newblock \bibinfo{journal}{Sci. Rep.} \bibinfo{volume}{12}
	(\bibinfo{year}{2022}) \bibinfo{pages}{560}.
	%Type = Article
	\bibitem[{Menezes and Barbalho(2023)}]{menezes_csf23}
	\bibinfo{author}{J.~Menezes}, \bibinfo{author}{R.~Barbalho},
	\newblock \bibinfo{title}{How multiple weak species jeopardise biodiversity in
		spatial rock-paper-scissors models},
	\newblock \bibinfo{journal}{Chaos, Solit. and Fract.} \bibinfo{volume}{169}
	(\bibinfo{year}{2023}) \bibinfo{pages}{113290}.
	%Type = Article
	\bibitem[{Avelino et~al.(2021)Avelino, \protect{de Oliveira}, and
		Trintin}]{avelino_epl21}
	\bibinfo{author}{P.~P. Avelino}, \bibinfo{author}{B.~F. \protect{de Oliveira}},
	\bibinfo{author}{R.~S. Trintin},
	\newblock \bibinfo{title}{Weak species in rock-paper-scissors models},
	\newblock \bibinfo{journal}{EPL} \bibinfo{volume}{134} (\bibinfo{year}{2021})
	\bibinfo{pages}{48001}.
	%Type = Article
	\bibitem[{Fu et~al.(2019)Fu, Guo, Cheng, Huang, and Chen}]{fu_mj_pa19}
	\bibinfo{author}{M.~Fu}, \bibinfo{author}{W.~Guo}, \bibinfo{author}{L.~Cheng},
	\bibinfo{author}{S.~Huang}, \bibinfo{author}{D.~Chen},
	\newblock \bibinfo{title}{History loyalty-based reward promotes cooperation in
		the spatial public goods game},
	\newblock \bibinfo{journal}{Physica A} \bibinfo{volume}{525}
	(\bibinfo{year}{2019}) \bibinfo{pages}{1323--1329}.
	%Type = Article
	\bibitem[{Zhong et~al.(2021)Zhong, Fan, and Di}]{zhong_xw_pa21}
	\bibinfo{author}{X.~Zhong}, \bibinfo{author}{Y.~Fan}, \bibinfo{author}{Z.~Di},
	\newblock \bibinfo{title}{The evolution of cooperation in public goods games on
		signed networks},
	\newblock \bibinfo{journal}{Physica A} \bibinfo{volume}{582}
	(\bibinfo{year}{2021}) \bibinfo{pages}{126217}.
	%Type = Article
	\bibitem[{Zheng et~al.(2022)Zheng, He, Ren, and Huang}]{zheng_jj_pa22}
	\bibinfo{author}{J.~Zheng}, \bibinfo{author}{Y.~He}, \bibinfo{author}{T.~Ren},
	\bibinfo{author}{Y.~Huang},
	\newblock \bibinfo{title}{Evolution of cooperation in public goods games with
		segregated networks and periodic invasion},
	\newblock \bibinfo{journal}{Physica A} \bibinfo{volume}{596}
	(\bibinfo{year}{2022}) \bibinfo{pages}{127101}.
	%Type = Article
	\bibitem[{Xie et~al.(2023)Xie, Liu, Wang, and Jiang}]{xie_k_csf23}
	\bibinfo{author}{K.~Xie}, \bibinfo{author}{X.~Liu}, \bibinfo{author}{H.~Wang},
	\bibinfo{author}{Y.~Jiang},
	\newblock \bibinfo{title}{Multi-heterogeneity public goods evolutionary game on
		lattice},
	\newblock \bibinfo{journal}{Chaos. Solit. and Fract.} \bibinfo{volume}{172}
	(\bibinfo{year}{2023}) \bibinfo{pages}{113562}.
	%Type = Article
	\bibitem[{Yang and Yang(2019)}]{yang_hx_pa19}
	\bibinfo{author}{H.-X. Yang}, \bibinfo{author}{J.~Yang},
	\newblock \bibinfo{title}{Reputation-based investment strategy promotes
		cooperation in public goods games},
	\newblock \bibinfo{journal}{Physica A} \bibinfo{volume}{523}
	(\bibinfo{year}{2019}) \bibinfo{pages}{886--893}.
	%Type = Article
	\bibitem[{Shen et~al.(2022)Shen, Yin, Kang, Zhang, and Wang}]{shen_y_pla22}
	\bibinfo{author}{Y.~Shen}, \bibinfo{author}{W.~Yin}, \bibinfo{author}{H.~Kang},
	\bibinfo{author}{H.~Zhang}, \bibinfo{author}{M.~Wang},
	\newblock \bibinfo{title}{High-reputation individuals exert greater influence
		on cooperation in spatial public goods game},
	\newblock \bibinfo{journal}{Phys. Lett. A} \bibinfo{volume}{428}
	(\bibinfo{year}{2022}) \bibinfo{pages}{127935}.
	%Type = Article
	\bibitem[{Xiao et~al.(2022)Xiao, Zhang, Li, Dai, and Yang}]{xiao_sl_epjb22}
	\bibinfo{author}{S.~Xiao}, \bibinfo{author}{L.~Zhang}, \bibinfo{author}{H.~Li},
	\bibinfo{author}{Q.~Dai}, \bibinfo{author}{J.~Yang},
	\newblock \bibinfo{title}{Environment-driven migration enhances cooperation in
		evolutionary public goods games},
	\newblock \bibinfo{journal}{Eur. Phys. J. B} \bibinfo{volume}{95}
	(\bibinfo{year}{2022}) \bibinfo{pages}{67}.
	%Type = Article
	\bibitem[{Kang et~al.(2021)Kang, Zhou, Shen, Sun, and Chen}]{kang_hw_pla21}
	\bibinfo{author}{H.~Kang}, \bibinfo{author}{X.~Zhou},
	\bibinfo{author}{Y.~Shen}, \bibinfo{author}{X.~Sun},
	\bibinfo{author}{Q.~Chen},
	\newblock \bibinfo{title}{Influencer propagation model promotes cooperation in
		spatial public goods game},
	\newblock \bibinfo{journal}{Phys. Lett. A} \bibinfo{volume}{417}
	(\bibinfo{year}{2021}) \bibinfo{pages}{127678}.
	%Type = Article
	\bibitem[{Deng and Zhang(2021)}]{deng_ys_pa21}
	\bibinfo{author}{Y.~Deng}, \bibinfo{author}{J.~Zhang},
	\newblock \bibinfo{title}{The role of the preferred neighbor with the expected
		payoff on cooperation in spatial public goods game under optimal strategy
		selection mechanism},
	\newblock \bibinfo{journal}{Physica A} \bibinfo{volume}{584}
	(\bibinfo{year}{2021}) \bibinfo{pages}{126363}.
	%Type = Article
	\bibitem[{Yang and Zhang(2021)}]{yang_lh_csf21}
	\bibinfo{author}{L.~Yang}, \bibinfo{author}{L.~Zhang},
	\newblock \bibinfo{title}{Environmental feedback in spatial public goods game},
	\newblock \bibinfo{journal}{Chaos, Solit. and Fract.} \bibinfo{volume}{142}
	(\bibinfo{year}{2021}) \bibinfo{pages}{110485}.
	%Type = Article
	\bibitem[{Feng and Liu(2023)}]{feng_s_csf23}
	\bibinfo{author}{S.~Feng}, \bibinfo{author}{X.~Liu},
	\newblock \bibinfo{title}{Effects of the limited incentive pool on cooperation
		evolution in public goods game},
	\newblock \bibinfo{journal}{Chaos, Solit. and Fract.} \bibinfo{volume}{169}
	(\bibinfo{year}{2023}) \bibinfo{pages}{113295}.
	%Type = Article
	\bibitem[{Liu et~al.(2021)Liu, Peng, Peng, Li, Chu, Li, and
		Liu}]{liu_jz_epjb21}
	\bibinfo{author}{J.~Liu}, \bibinfo{author}{M.~Peng}, \bibinfo{author}{Y.~Peng},
	\bibinfo{author}{Y.~Li}, \bibinfo{author}{C.~Chu}, \bibinfo{author}{X.~Li},
	\bibinfo{author}{Q.~Liu},
	\newblock \bibinfo{title}{Effects of inequality on a spatial evolutionary
		public goods game},
	\newblock \bibinfo{journal}{Eur. Phys. J. B} \bibinfo{volume}{94}
	(\bibinfo{year}{2021}) \bibinfo{pages}{167}.
	%Type = Article
	\bibitem[{Huang et~al.(2023)Huang, Ren, Zheng, Liu, and Zhang}]{huang_yc_amc23}
	\bibinfo{author}{Y.~Huang}, \bibinfo{author}{T.~Ren},
	\bibinfo{author}{J.~Zheng}, \bibinfo{author}{W.~Liu},
	\bibinfo{author}{M.~Zhang},
	\newblock \bibinfo{title}{Evolution of cooperation in public goods games with
		dynamic resource allocation: A fairness preference perspective},
	\newblock \bibinfo{journal}{Appl. Math. and Comput.} \bibinfo{volume}{445}
	(\bibinfo{year}{2023}) \bibinfo{pages}{127844}.
	%Type = Article
	\bibitem[{Yu et~al.(2022)Yu, Wang, and He}]{yu_fy_csf22}
	\bibinfo{author}{F.~Yu}, \bibinfo{author}{J.~Wang}, \bibinfo{author}{J.~He},
	\newblock \bibinfo{title}{Inequal dependence on members stabilizes cooperation
		in spatial public goods game},
	\newblock \bibinfo{journal}{Chaos, Solit. and Fract.} \bibinfo{volume}{165}
	(\bibinfo{year}{2022}) \bibinfo{pages}{112755}.
	%Type = Article
	\bibitem[{He et~al.(2022)He, Wang, Yu, Chen, and Ji}]{he_jl_pla22}
	\bibinfo{author}{J.~He}, \bibinfo{author}{J.~Wang}, \bibinfo{author}{F.~Yu},
	\bibinfo{author}{W.~Chen}, \bibinfo{author}{Y.~Ji},
	\newblock \bibinfo{title}{The interplay between reputation and heterogeneous
		investment enhances cooperation in spatial public goods game},
	\newblock \bibinfo{journal}{Phys. Lett. A} \bibinfo{volume}{442}
	(\bibinfo{year}{2022}) \bibinfo{pages}{128182}.
	%Type = Article
	\bibitem[{Ding et~al.(2023)Ding, Wang, Zhao, Gu, and Wang}]{ding_r_csf23}
	\bibinfo{author}{R.~Ding}, \bibinfo{author}{X.~Wang},
	\bibinfo{author}{J.~Zhao}, \bibinfo{author}{C.~Gu},
	\bibinfo{author}{T.~Wang},
	\newblock \bibinfo{title}{The evolution of cooperation in spatial public goods
		games under a risk-transfer mechanism},
	\newblock \bibinfo{journal}{Chaos, Solit. and Fract.} \bibinfo{volume}{169}
	(\bibinfo{year}{2023}) \bibinfo{pages}{113236}.
	%Type = Article
	\bibitem[{Szolnoki et~al.(2020)Szolnoki, \protect{de Oliveira}, and
		Bazeia}]{szolnoki_epl20}
	\bibinfo{author}{A.~Szolnoki}, \bibinfo{author}{B.~F. \protect{de Oliveira}},
	\bibinfo{author}{D.~Bazeia},
	\newblock \bibinfo{title}{Pattern formations driven by cyclic interactions: A
		brief review of recent developments},
	\newblock \bibinfo{journal}{EPL} \bibinfo{volume}{131} (\bibinfo{year}{2020})
	\bibinfo{pages}{68001}.
	%Type = Article
	\bibitem[{Park and Jang(2019)}]{park_c19b}
	\bibinfo{author}{J.~Park}, \bibinfo{author}{B.~Jang},
	\newblock \bibinfo{title}{Robust coexistence with alternative competition
		strategy in the spatial cyclic game of five species},
	\newblock \bibinfo{journal}{Chaos} \bibinfo{volume}{29} (\bibinfo{year}{2019})
	\bibinfo{pages}{051105}.
	%Type = Article
	\bibitem[{Yang and Park(2023)}]{yang_csf23}
	\bibinfo{author}{R.~K. Yang}, \bibinfo{author}{J.~Park},
	\newblock \bibinfo{title}{Evolutionary dynamics in the cyclic competition
		system of seven species: Common cascading dynamics in biodiversity},
	\newblock \bibinfo{journal}{Chaos, Solit. and Fract.} \bibinfo{volume}{175}
	(\bibinfo{year}{2023}) \bibinfo{pages}{113949}.
	
\end{thebibliography}

\end{document}